\documentclass[pra,twocolumn,amsmath,amssymb,floatfi,showpacs]{revtex4}
\usepackage{graphicx}

\begin{document}
\title{\textbf{Nonlinear theory of laser-induced dipolar interactions in arbitrary geometry}}
\author{Ephraim Shahmoon}
\author{Gershon Kurizki}
\affiliation{Department of Chemical Physics, Weizmann Institute of Science, Rehovot, 76100, Israel}
\date{\today}

\begin{abstract}
Polarizable dipoles, such as atoms, molecules or nanoparticles, subject to laser radiation, may attract or repel each other. We derive a general formalism in which such laser-induced dipole-dipole interactions (LIDDI) in any geometry and for any laser strength are described in terms of the resonant dipole-dipole interaction (RDDI) between dipoles dressed by the laser. This approach provides a simple route towards the analysis of LIDDI in a general geometry. Our general results reveal LIDDI effects due to nonlinear dipolar response to the laser, previously unaccounted for. The origin of these nonlinear effects is discussed. Our general formalism is illustrated for LIDDI between atoms in a cavity.
\end{abstract}

 \maketitle

\section{Introduction}

Electromagnetic fields are the main tool in the control and manipulation of the motion of polarizable dipoles, such as atoms and molecules, which are key objects in physics and chemistry. Optical traps \cite{PS} and laser cooling \cite{CCTn} are widely used to induce external mechanical forces on individual dipoles, whereas interactions between atoms are often tuned with the help of external magnetic or electric fields \cite{PS,revG}. In this work, we address laser-induced dipole-dipole interactions (LIDDI) by establishing their relation to resonant inter-dipolar interaction (excitation exchange) in any confined geometry.

Polarizable dipoles subject to static electric fields interact via the electrostatic dipole-dipole interaction that scales in free space as $1/r^3$, $r$ being the inter-dipolar distance. A dynamic analog is obtained when a laser field, far-detuned from the dipolar resonant frequency, illuminates the (induced) dipoles. In the retarded regime, where $k_L r\gg 1$ with $k_L$ the laser wavenumber, the resulting LIDDI scales like $\cos (k_L r)/r$, whereas in the non-retarded quasistatic case, where $k_L r\ll 1$, the electrostatic scaling $1/r^3$ is restored \cite{MQED,THI,SAL}. The retarded long-range LIDDI was shown to be associated with peculiar many-body effects, such as self trapping \cite{OD1}, density modulations \cite{GIO} and the existence of a roton-like collective excitations \cite{OD2}, in an atomic Bose-Einstein condensate.

Recently, the possibility to control and shape the space-dependence of the inter-dipolar LIDDI potential was considered. One option is to tune the laser parameters \cite{LEM} or the radiation spectrum \cite{AND1}. Another, more recent approach, is to consider dipoles coupled to structures that support confined photon modes \cite{CHA,LIDDI}. Once illuminated by off-resonant light and virtually excited, these dipoles interact via confined virtual photons whose spatial-mode structure determines the resulting LIDDI space-dependence. For example, in the case of many atoms that are trapped in the vicinity of an optical fiber \cite{KIM} and free to move along its axis as in \cite{RAU}, the fiber-mediated LIDDI can effectively become one-dimensional (1d), such that it extends to any range and the atoms may self-organize \cite{CHA}. When the fiber incorporates a Bragg grating (1d photonic crystal), the relaxation dynamics of the resulting many-atom system was shown to reveal the inequivalence of statistical ensembles typical of non-additive systems \cite{LIDDI}. Self organization and dynamics of laser-illuminated atoms inside a cavity were also studied theoretically and experimentally \cite{RDE}.

A similar situation arises when considering the spatial dependence of the dispersive interaction between the internal degrees of freedom of the dipoles, the so-called resonant dipole-dipole interaction (RDDI), where a dipole, e.g. an atom initially in the excited state, periodically exchanges its excitation with another dipole, e.g. an atom initially in the ground state \cite{MQED,LEH}. RDDI is mediated by the exchange of virtual photons between the atoms, hence its space-dependence is determined by the spatial structure/propagation of the mediating photon modes. In free space and for $r\ll \lambda$, $\lambda$ being the typical dipolar transition wavelength, the RDDI exchange frequency scales as  $1/r^3$ whereas in the retarded regime $r\gg \lambda$ it takes the form $ \cos (2\pi r/\lambda)/r$. Such RDDI retardation effects are of interest for quantum information processing \cite{GER}. Studies of geometries where RDDI is mediated by confined photon modes include, for example, RDDI via surface-plasmon-polariton and coplanar waveguide modes in one dimension \cite{SPA,SAN} or in geometries that create cutoffs or bandgaps in the photonic mode spectrum \cite{KUR,SEK,LAW}, where spontaneous emission is absent and enhanced long-range RDDI may become dominant \cite{RDDI}.

The extensive experimental progress in the ability to couple atoms/dipoles to various confined photonic structures \cite{KIM,RAU,GAE,HAR,LTT,TRA}, makes the discussion of LIDDI in such structures ever more relevant. In this study we derive a general formalism for LIDDI in \emph{any geometry} by relating it to the underlying RDDI process, thus providing a simple and intuitive recipe for calculating LIDDI through the less complex RDDI. We also show that the same holds for the relation between scattering of laser photons and spontaneous emission, respectively.

Our formalism has several appealing features: (1) It is fully quantum mechanical, with the laser radiation taken to be in a coherent state (in contrast to the less realistic number state \cite{MQED}). (2) It interprets \emph{LIDDI as RDDI between dressed dipoles}, thus allowing to account for LIDDI in strong radiation fields. This considerably simplifies the calculation of LIDDI in a general geometry to merely calculating the RDDI in that geometry, thereby avoiding the need for the rather cumbersome fourth-order perturbative QED calculation which contains 24 diagrams \cite{MQED,SAL}. (3) It enables the analysis of \emph{nonlinear effects} in LIDDI, resulting from the nonlinear response of the dipoles to the laser, unaccounted for previously. (4) For the case of far-detuned laser field, it reveals a simple relation between the pairwise RDDI energy $\hbar \Delta_{12}$ and the LIDDI potential $U\propto (\Omega^2/\delta^2)\hbar \Delta_{12}$, where $\Omega$ is the laser Rabi frequency and $\delta$ the dipole-laser detuning. This provides an intuitive picture for LIDDI: The dipole/atom is virtually excited by the off-resonant laser with probability $\Omega^2/\delta^2$ and, once excited, interacts with the other dipole/atom via RDDI.

The paper is organized as the following. Section II is dedicated to the derivation of the general formalism that relates LIDDI and scattering to their underlying RDDI and spontaneous emission, respectively, arriving at our central, general result for the LIDDI potential, Eq. (\ref{LIDDI1}). The large-detuning limit is discussed in Section III, whereas the application of our formalism to the analysis of LIDDI in a cavity is illustrated in Section IV.
In Section V we put forward intuitive and simple arguments that explain the origin of LIDDI effects caused by nonlinear dipole response to the laser. Our conclusions are presented in Section VI.

\section{General LIDDI-RDDI formalism}
We adopt a two-level atom model for the polarizable dipole. Thus, while still capturing the essence of nonlinear response of the dipoles, the discussion is kept simple and intuitive. The two-level atom approximation is justified for the typical situation where a laser frequency is close to a specific dipole-allowed transition of a system with discrete energy levels, such as atoms, molecules, quantum dots or even superconducting qubits. The atoms, with excited state  $|e\rangle$ and ground state $|g\rangle$ interact with a laser ($H_A$ below). Considering a laser mode in a coherent state, an \emph{exact} transformation due to Mollow \cite{MOL1} allows to treat the interacting system of atoms + single field mode in a coherent state (laser) + vacuum in the rest of the modes, as a system of atoms + single-mode external field (laser) + vacuum in all field modes \cite{MOL1,CCT}. The atoms are coupled to the vacuum modes $\{k\}$ in the considered geometry ($H_V$ below) via dipole couplings ($H_{AV}$ below) which ultimately lead to the inter-atomic interaction.
The Hamiltonian $H=H_A+H_V+H_{AV}$ reads
\begin{eqnarray}
H_A&=&\sum_{\nu=1}^2\frac{1}{2}\hbar\left[\omega_{e} \hat{\sigma}_{\nu}^{z}+\left(e^{i(\mathbf{k}_L\cdot\mathbf{r}_{\nu}-\omega_L t)}\Omega \hat{\sigma}^{+}_{\nu}+\mathrm{h.c.}\right)\right],
\nonumber\\
H_V&=&\sum_k\hbar\omega_k\hat{a}^{\dag}_k\hat{a}_k,
\nonumber \\
H_{AV}&=&\sum_{\nu=1}^2\sum_k\hbar\left[i g_{k\nu} \hat{a}_k  - i g^{\ast}_{k\nu} \hat{a}^{\dag}_k \right]\left[\hat{\sigma}_{\nu}^{-}+\hat{\sigma}_{\nu}^{+}\right].
\nonumber \\
\label{H}
\end{eqnarray}
Here $\hat{a}_k$ is the electromagnetic vacuum mode $k$ lowering operator, whereas $\omega_L$ and $\mathbf{k}_L$ are the laser frequency and wavevector with Rabi frequency $\Omega=E_L\mathbf{e}_L\cdot\mathbf{d}/\hbar$, $E_L$ and $\mathbf{e}_L$ being its electric field amplitude and polarization vector, respectively, and $\mathbf{d}$ denoting the atomic dipole matrix element. In order to find the pairwise-LIDDI potential it is enough to focus on a single pair of atoms with indices $\nu=1,2$ and corresponding atomic operators $\hat{\sigma}^-=|g\rangle \langle e|=(\hat{\sigma}^+)^{\dag}$, $\hat{\sigma}^z=|e\rangle\langle e|-|g\rangle\langle g|$. The atom-vacuum dipolar couplings are $g_{k\nu}=\sqrt{\frac{\omega_k}{2\epsilon_0\hbar}}\mathbf{d}\cdot\mathbf{u}_k(\mathbf{r}_{\nu})$,
$\mathbf{r}_{\nu}$ being the location of atom $\nu$, and $\mathbf{u}_k(\mathbf{r})$ the spatial function of mode $k$.    Moving to the interaction picture with respect to (w.r.t) $H_A+H_V$ the Hamiltonian becomes
\begin{equation}
H_I(t)=\sum_{\nu=1}^2 \sum_k\hbar \left[ig_{k\nu} \hat{a}_k \tilde{\sigma}^+_{\nu}(t)e^{-i\omega_k t}+\mathrm{h.c.}\right].
\label{HI1}
\end{equation}
Here the rotating-wave approximation $t\gg 1/\omega_e,1/\omega_L$ is taken, and the interaction-picture atomic operators are $\tilde{\sigma}^{\pm}_{\nu}(t)=\hat{U}_A^{\dag}(t)\hat{\sigma}^{\pm}_{\nu}\hat{U}_A(t)$ with $\hat{U}_A=\mathcal{T}e^{-\frac{i}{\hbar}\int_0^tdt'H_A(t')}$, where $\mathcal{T}$ denotes time ordering.

\subsection{Atom-laser system}
We are now interested in finding the interaction-picture operator $\tilde{\sigma}^{\pm}(t)$ of the coupled atom-laser system. This is done for each atom $\nu=1,2$ separately, hence the index $\nu$ is suppressed here. The operator $\tilde{\sigma}^{\pm}(t)$ can  be viewed as a Heisenberg-picture operator w.r.t the Hamiltonian $H_A$ from Eq. (\ref{H}). Then, dividing $H_A$ into $H_A-H_L$ and $H_L=(1/2)\hbar\omega_L \hat{\sigma}^z$, we move to the so-called laser-rotated frame, which is an interaction picture  w.r.t $H_L$, where the Hamiltonian $H_A$ becomes
\begin{equation}
H'_A=-\frac{1}{2}\hbar \delta \hat{\sigma}^z+\hbar \frac{1}{2}(\tilde{\Omega}\hat{\sigma}^+ + \mathrm{h.c.}),
\label{HA}
\end{equation}
with $\tilde{\Omega}=\Omega e^{i\mathbf{k}_L\cdot\mathbf{r}}$ and where $\delta=\omega_L-\omega_e$ is the atom-laser detuning. The eigenstates $|\pm\rangle$ and eigenvalues $\hbar\varepsilon_{\pm}$ of $H'_A$ are found to be (see also \cite{CCT,BAR})
\begin{eqnarray}
|+\rangle&=&\frac{(\delta+\bar{\Omega})}{\sqrt{2\bar{\Omega}(\bar{\Omega}+\delta)}}|g\rangle + \frac{\tilde{\Omega}}{\sqrt{2\bar{\Omega}(\bar{\Omega}+\delta)}}|e\rangle,
\nonumber \\
|-\rangle&=&\frac{(\delta-\bar{\Omega})}{\sqrt{2\bar{\Omega}(\bar{\Omega}-\delta)}}|g\rangle + \frac{\tilde{\Omega}}{\sqrt{2\bar{\Omega}(\bar{\Omega}-\delta)}}|e\rangle,
\nonumber \\
\varepsilon_{\pm}&=&\pm\frac{1}{2}\bar{\Omega},
\label{tran}
\end{eqnarray}
where $\bar{\Omega}=\sqrt{|\Omega|^2+\delta^2}$ is the effective Rabi frequency.
Considering the relation between operators in the Heisenberg and interaction pictures we find
\begin{equation}
\tilde{\sigma}^{\pm}(t)=U_A^{'\dag}U_L^{\dag}\hat{\sigma}^{\pm}U_L U'_A =e^{\pm i\omega_L t}U_A^{'\dag}\hat{\sigma}^{\pm}U'_A,
\label{UA}
\end{equation}
where $U'_A=e^{-(i/\hbar)H'_A t}$ and $U_L=e^{-(i/\hbar)H_L t}$, such that $U'_A|\pm\rangle=e^{-i\varepsilon_{\pm}t}|\pm\rangle$. Then, recalling $\hat{\sigma}^+=|e\rangle \langle g|$, $\hat{\sigma}^-=|g\rangle \langle e|$, we can use the transformation (\ref{tran}) to express the operators $\hat{\sigma}^{\pm}$ in terms of the $|\pm\rangle$-dressed-basis projection operators and easily apply $U'_A$ on these operators, obtaining
\begin{eqnarray}
\tilde{\sigma}^{+}_{\nu}(t)&=&e^{i(\omega_L-\bar{\Omega})t}\frac{\delta+\bar{\Omega}}{2\bar{\Omega}}e^{-i\mathbf{k}_L\cdot \mathbf{r}_{\nu}}\hat{S}^{-}_{\nu}+
e^{i\omega_Lt}\frac{|\Omega|}{2\bar{\Omega}}e^{-i\mathbf{k}_L\cdot \mathbf{r}_{\nu}}\hat{S}^{z}_{\nu}
\nonumber \\
&+&
e^{i(\omega_L+\bar{\Omega})t}\frac{\delta-\bar{\Omega}}{2\bar{\Omega}}e^{-i\mathbf{k}_L\cdot \mathbf{r}_{\nu}}\hat{S}^{+}_{\nu},
\label{sig1}
\end{eqnarray}
where $\hat{S}^{\pm}=|\pm\rangle \langle \mp|$ and $\hat{S}^{z}=|+\rangle \langle +|-|-\rangle \langle -|$ are the spin operators in the dressed-basis. This result can be written in a more compact form by absorbing the time-independent coefficients into the operators, as
\begin{eqnarray}
&&\tilde{\sigma}_{\nu}^+(t)=\sum_{i=\pm,z}e^{i\omega_i t}\tilde{S}_{i\nu},
\nonumber \\
&& \tilde{S}_{\pm\nu}=\frac{\delta\mp\bar{\Omega}}{2\bar{\Omega}}e^{-i\mathbf{k}_L\cdot \mathbf{r}_{\nu}}\hat{S}^{\pm}_{\nu},\quad
\tilde{S}_{z\nu}=\frac{|\Omega|}{2\bar{\Omega}}e^{-i\mathbf{k}_L\cdot \mathbf{r}_{\nu}}\hat{S}^{z}_{\nu},
\nonumber\\
&&\omega_{\pm}=\omega_L{\pm}\bar{\Omega}, \quad \omega_z=\omega_L.
\label{sig2}
\end{eqnarray}
Eqs. (\ref{sig1}) and (\ref{sig2}) describe the dynamics of the atom operator $\sigma^+$, dressed by the laser, as composed of a linear combination of the three spin operators in the dressed-basis $i=\pm,z$, each oscillating in time with a distinct frequency $\omega_i$, hence allowing all possible transitions in this basis.

\subsection{Master-equation approach}
Inserting $\tilde{\sigma}_{\nu}^{+}(t)$ from Eq. (\ref{sig2}) into Eq. (\ref{HI1}), we obtain the interaction picture Hamiltonian in the form
\begin{equation}
H_I(t)=\sum_{\nu=1}^2\sum_{i=\pm,z} \sum_k\hbar \left[ig_{k\nu} \hat{a}_k \tilde{S}_{i\nu}e^{-i(\omega_k-\omega_i) t}+\mathrm{h.c.}\right].
\label{HI2}
\end{equation}
This Hamiltonian describes the interaction of atoms, dressed by the laser (operators $\tilde{S}_{i\nu}$), with the electromagnetic vacuum (operators $\hat{a}_k$). We are interested to use lowest order perturbation theory in order to find the effective interaction between the dressed atoms, mediated by the vacuum, similar to the case of RDDI \cite{LEH,RDDI}. Equivalently, this can be performed by the derivation of the master equation  for the density operator of the two-atom system $\rho(t)$ when it interacts with the vacuum reservoir in a stationary vacuum state $\rho_V=|0\rangle\langle0|$. To lowest order (Born approximation) the master equation is given by \cite{CAR}
\begin{equation}
\dot{\rho}=-\frac{1}{\hbar^2}\int_0^t dt' \mathrm{tr}_V \left\{\left[H_I(t),[H_I(t'),\rho(t')\rho_V]\right]\right\}.
\label{ME1}
\end{equation}
Inserting the Hamiltonian from Eq. (\ref{HI2}) into Eq. (\ref{ME1}) and using standard methods (see Appendix A), we obtain the Markovian master equation
\begin{eqnarray}
\dot{\rho}&=&-\frac{i}{\hbar}[H_{DD},\rho]+\sum_{\nu,\nu'=1}^2\sum_{i=\pm,z}\left[\Gamma_{\nu \nu'}^i\tilde{S}_{i\nu}^{\dag}\rho\tilde{S}_{i\nu'}
\right.
\nonumber \\
&& \left.
-\frac{1}{2}\Gamma_{\nu \nu'}^i\left(\tilde{S}_{i\nu}\tilde{S}_{i\nu'}^{\dag}\rho+\rho\tilde{S}_{i\nu}\tilde{S}_{i\nu'}^{\dag}\right)\right],
\label{ME2}
\end{eqnarray}
with the effective dipole-dipole Hamiltonian
\begin{equation}
H_{DD}=-\frac{1}{2}\sum_{\nu,\nu'=1}^2\sum_{i=\pm,z} \left(\hbar \Delta_{\nu\nu'}^i\tilde{S}_{i\nu}\tilde{S}_{i\nu'}^{\dag}+\mathrm{h.c.}\right).
\label{HDD}
\end{equation}
The dispersive, virtual-photon-mediated, RDDI energy $\Delta_{\nu\nu'}^i$ and the corresponding cooperative emission rate $\Gamma_{\nu\nu'}^i$, are related by Kramers-Kronig relation and given by
\begin{eqnarray}
&&\Delta_{\nu\nu'}^i=\Delta_{\nu\nu'}(\omega_i)=\mathrm{P}\int_0^{\infty} d\omega\frac{ G_{\nu\nu'}(\omega)}{\omega-\omega_i},
\nonumber \\
&&\Gamma_{\nu\nu'}^i=\Gamma_{\nu\nu'}(\omega_i)=2\pi G_{\nu\nu'}(\omega_i),
\label{D}
\end{eqnarray}
where $G_{\nu\nu'}(\omega)$ is the vacuum-reservoir two-point (autocorrelation) spectrum defined in the continuum limit $\sum_k g_{k\nu}g_{k\nu'}^{\ast}\longrightarrow \int d \omega G_{\nu \nu'}(\omega)$. The geometry dependence of the LIDDI and related effects discussed below is thus encoded in the function $G_{\nu\nu'}(\omega)$, since this function depends on the dipole couplings $g_{k\nu}$, which in turn depend on the spatial photon modes of the confined geometry $\mathbf{u}_k(\mathbf{r}_{\nu})$.

Two important assumptions regarding the temporal resolution of interest (coarse-graining time) $T$ were made in the derivation of Eqs. (\ref{ME2}), (\ref{HDD}) and (\ref{D}) (Appendix A): (1) $T\gg 1/|\omega_i-\omega_j|, \forall i\neq j$, i.e. $T\gg 1/\sqrt{|\Omega|^2+\delta^2}$. This allows to take terms oscillating as $e^{i(\omega_i-\omega_j)t}$ to be Kronecker deltas $\delta_{ij}$. (2) $T\gg \tau_c$, where $\tau_c$ is the correlation time of the reservoir, namely $1/\tau_c$ is the width of the functions $\Delta_{\nu\nu'}(\omega),\Gamma_{\nu\nu'}(\omega)$ around $\omega_i$. This allows to take the Markov approximation and obtain a master equation which is local in time (see Appendix A and Refs. \cite{RDDI,CAR}).

\subsection{RDDI and cooperative emission}
\subsubsection{Bare atom}
In the absence of the external laser light, i.e. taking the limit $\Omega\rightarrow 0$ in Eqs. (\ref{sig2}) and (\ref{tran}), we obtain $\tilde{S}_{+\nu}=0$, $\tilde{S}_{z\nu}=0$, $\tilde{S}_{-\nu}=\hat{\sigma}^+_{\nu}$ and $\omega_-=\omega_e$, such that the effective dipole-dipole Hamiltonian, Eq. (\ref{HDD}), reduces to that of RDDI,
\begin{equation}
H_{DD}=-\frac{1}{2}\sum_{\nu\nu'}\left(\hbar\Delta^-_{\nu\nu'}\hat{\sigma}_{\nu}^+\hat{\sigma}_{\nu'}^- +\mathrm{h.c.}  \right).
\label{RDDI}
\end{equation}
Then, once one of the atoms happens to be excited, it coherently and reversibly exchanges its excitation with the other, initially unexcited, atom. The complementary incoherent and radiative cooperative process, namely that of irreversible excitation exchange between the atoms via the emission of a real photon at the atomic frequency $\omega_e$ \cite{LEH,DIC}, is described in the non-Hamiltonian part of the master equation (\ref{ME2}) by terms proportional to $\Gamma_{\nu\nu'}^-$, e.g. $\Gamma_{\nu\nu'}^- \hat{\sigma}_{\nu}^+\hat{\sigma}_{\nu'}^- \rho$. These bare-atom processes are depicted by the diagram in Fig. 1(a), where the atoms exchange their excitation via a photon at frequency $\omega$. In the radiative process, the photon is real and hence has the frequency of the atomic transition $\omega=\omega_e$, whereas the dispersive effect involves the summation over all diagrams with virtual photons $\omega$ around $\omega_e$, as can be seen by the sampling of the functions $\Gamma_{\nu\nu'}(\omega)$ and  $\Delta_{\nu\nu'}(\omega)$ in Eq. (\ref{D}), at $\omega_-=\omega_e$ for $\Gamma^-_{\nu\nu'}$ and $\Delta^-_{\nu\nu'}$, respectively. Both RDDI and its irreversible counterpart rely on the fact that one of the atoms is excited, and are absent when both atoms are in their ground states. As we shall see below, the essence of LIDDI revealed by our formalism is the possibility to excite the atoms by the laser, thus allowing them to interact via RDDI.

\begin{figure}
\begin{center}
\includegraphics[scale=0.4]{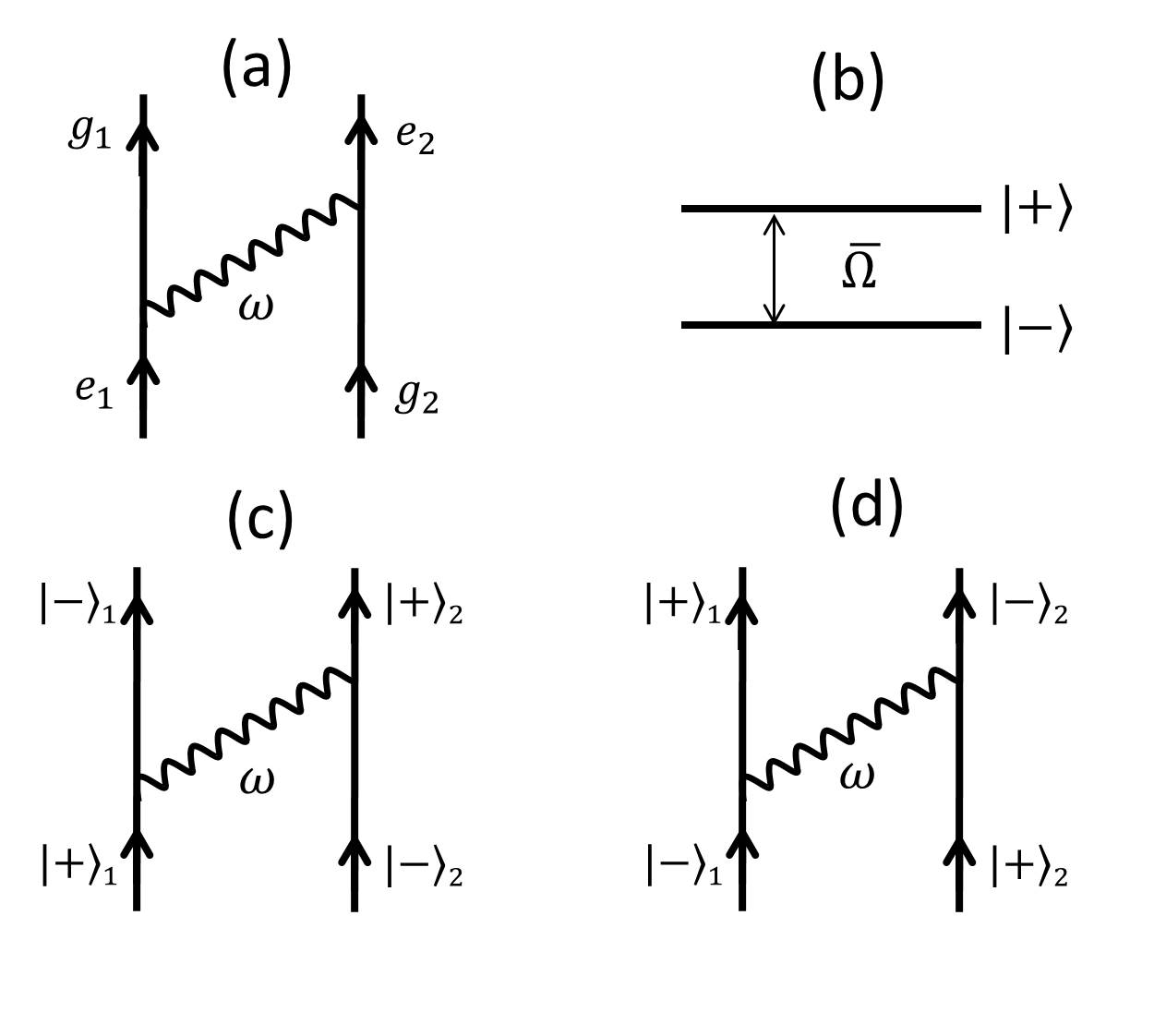}
\caption{\small{
(a) Interaction of bare atoms: Atom 1, initially in the excited state $|e_1\rangle$, emits a photon at frequency $\omega$ to atom 2, initially in the ground state $g_2\rangle$. As a result, the atoms flip their states to $g_1\rangle$ and $|e_2\rangle$, respectively. (b) Level diagram of a dressed atom, consisting of the dressed states $|\pm\rangle$, in a laser-rotated frame (by frequency $\omega_L$). (c) Interaction of dressed atoms by a $|+\rangle\rightarrow |-\rangle$ transition: $|+\rangle$ and $|-\rangle$ are equivalent to $|e\rangle$ and $|g\rangle$, respectively, in the bare-atom case. (d) Interaction of dressed atoms by a $|-\rangle\rightarrow |+\rangle$ transition: As in (c), except that here, the atom that emits the photon (atom 1) is initially in the $|-\rangle$ state, whereas the absorbing atom 2 is initially in the state $|+\rangle$.
}}
\end{center}
\end{figure}

\subsubsection{Dressed atom}
In the more general case, where the laser is turned on, we write the effective dipole-dipole Hamiltonian (\ref{HDD}) explicitly using Eq. (\ref{sig2}), as
\begin{eqnarray}
H_{DD}&=&-\hbar\Delta^z_{12}\frac{|\Omega|^2}{2\bar{\Omega}^2}\cos(\mathbf{k}_L\cdot\mathbf{r}_{12})\hat{S}^z_1\hat{S}^z_2
\nonumber \\
&-&\hbar\Delta^+_{12}\frac{(\delta-\bar{\Omega})^2}{4\bar{\Omega}^2}\left[e^{i\mathbf{k}_L\cdot\mathbf{r}_{12}}\hat{S}^-_1\hat{S}^+_2+e^{-i\mathbf{k}_L\cdot\mathbf{r}_{12}}\hat{S}^+_1\hat{S}^-_2\right],
\nonumber \\
&-&\hbar\Delta^-_{12}\frac{(\delta+\bar{\Omega})^2}{4\bar{\Omega}^2}\left[e^{-i\mathbf{k}_L\cdot\mathbf{r}_{12}}\hat{S}^-_1\hat{S}^+_2+e^{i\mathbf{k}_L\cdot\mathbf{r}_{12}}\hat{S}^+_1\hat{S}^-_2\right],
\nonumber \\
\label{HDD_D}
\end{eqnarray}
where $\mathbf{r}_{12}=\mathbf{r}_{1}-\mathbf{r}_{2}$. We also note that here the single-atom $\Delta^i_{\nu\nu}$ ($\nu=\nu'$) terms are not considered, since they are not important for the interaction between the dipoles, and that we have assumed the typical situation where $\Delta^i_{12}=\Delta^{i\ast}_{21}$ is real. The complementary dissipative effects include terms of similar form, e.g. with $\propto i\Gamma^{\pm}_{12}(\delta\mp\bar{\Omega})^2/(4\bar{\Omega}^2)$ and $\propto i\Gamma^{z}_{12}|\Omega|^2/(4\bar{\Omega}^2)$ replacing the corresponding $\Delta^i_{12}$ terms in (\ref{HDD_D}).

In analogy to the bare-atom case, the above Hamiltonian and its dissipative counterparts describe the RDDI and cooperative emission, respectively, between two dressed atoms, whose level scheme, consisting of two levels separated by an energy $\bar{\Omega}$ [Eq. (\ref{tran})], is presented in Fig. 1(b). However, as opposed to the bare-atom case, where only one exchange process exists [see Eq. (\ref{RDDI}) and Fig. 1(a)], here several processes can occur:
\\
(i) The first line in Eq. (\ref{HDD_D}) describes a process where the internal states of the dressed atoms do not change, giving rise to a cooperative energy shift to the dressed states.
\\
(ii) The process described in the second line, shown in Fig. 1(c), is analogous to the bare-atom RDDI in Eq. (\ref{RDDI}) and Fig. 1(a), where the dressed states $|+\rangle$ and $|-\rangle$ take the role of the excited and ground states $|e\rangle$ and $|g\rangle$, respectively. The relevant atomic transition is then $|+\rangle\rightarrow |-\rangle$, which in Fig. 1(b) is shown to be resonant with $\bar{\Omega}$. However, since the dressed-states (\ref{tran}) are written in a frame rotating with frequency $\omega_L$, the actual transition frequency is $\omega_+=\omega_L+\bar{\Omega}$. Then, a real-photon, dissipative effect involves a photon exchange with frequency $\omega=\omega_+$, whereas the dispersive RDDI results from the equivalent virtual-photon effect, just as seen by the sampling of $\Gamma_{12}(\omega)$ and $\Delta_{12}(\omega)$ at $\omega=\omega_+$ in Eq. (\ref{D}) for $\Gamma^{+}_{12}$ and $\Delta^{+}_{12}$, respectively.
\\
(iii) The last process, described by the third line of (\ref{HDD_D}) is that of the atomic transition $|-\rangle\rightarrow |+\rangle$, shown in Fig. 1(d). Namely, atom 1, initially in state $|-\rangle$, transfers a photon to atom 2 which is initially in state $|+\rangle$, and as a result both of them flip their states. The emission of a photon from the state $|-\rangle$ and its absorption by the state $|+\rangle$ underlying this process, is enabled by the fact that the dressed state $|-\rangle$ contains a component of the excited state and likewise the  dressed state $|+\rangle$ contains a ground-state component. The transition frequency in the dressed-state basis is then $-\bar{\Omega}$, so that the actual frequency is $\omega_-=\omega_L-\bar{\Omega}$.

The three processes surveyed above, describe the interaction between the atoms mediated by photon modes at frequencies $\omega_{\pm}=\omega_L\pm\bar{\Omega}$ and $\omega_z=\omega_L$, respectively, in analogy with the three spectral components of the well-known Mollow-triplet of resonance fluorescence \cite{MOL2}. The distinction between these three processes and their physical meaning are further discussed in Sec. V C below.

\subsection{LIDDI potential and scattering rate}

We now present a central insight of this paper, namely, that LIDDI (and scattering) originate from RDDI (and emission) of dressed atoms.
The average interaction energy between the dressed atoms, interpreted as the LIDDI potential, is obtained by averaging quantum mechanically over the Hamiltonian $H_{DD}$, Eq. (\ref{HDD_D}),
\begin{equation}
U(\mathbf{r}_{12})=\langle H_{DD} \rangle=\mathrm{tr}[\rho H_{DD}],
\label{LIDDI}
\end{equation}
where $\rho$ is the density matrix of the two-atom state. We note that the Hamiltonian $H_{DD}$ is assumed to be time-independent in this formalism. In fact, $H_{DD}$ depends on the external degrees of freedom of the atoms, namely their positions $\mathbf{r}_{1,2}$, and hence becomes time-dependent due to their dynamics which is driven by the LIDDI potential $U(\mathbf{r}_{12})$. However, assuming this dynamics is slow enough, namely, $\sim\dot{\mathbf{r}}_{12}/\mathbf{r}_{12}$ is much smaller than any relevant energy scale such as $\bar{\Omega}$, a description using the potential $U(\mathbf{r}_{12})$ from Eq. (\ref{LIDDI}) can be viewed as a valid adiabatic approximation.

\subsubsection{State-dependent potential}
Eqs. (\ref{HDD_D}) and (\ref{LIDDI}) allow us to write the LIDDI potential for a general two-atom density operator $\rho$. Using the two-atom basis,
\begin{equation}
|1\rangle=|+_1+_2\rangle, \,\, |2\rangle=|+_1 -_2\rangle, \,\, |3\rangle=|-_1 +_2\rangle, \,\, |4\rangle=|-_1 -_2\rangle,
\label{base}
\end{equation}
and denoting the corresponding density matrix elements $\langle n |\rho |m\rangle =\rho_{nm}$, we find $\mathrm{tr}[\rho \hat{S}^z_1\hat{S}^z_2]=1-2(\rho_{22}+\rho_{33})$ and $\mathrm{tr}[\rho \hat{S}^-_1\hat{S}^+_2]=\rho_{23}$. The LIDDI potential between a pair of atoms is then,
\begin{eqnarray}
U(\mathbf{r}_{12})&=&U^z(\mathbf{r}_{12})+U^+(\mathbf{r}_{12})+U^-(\mathbf{r}_{12}),
\nonumber \\
U^z(\mathbf{r}_{12})&=&-\hbar\Delta^z_{12}\frac{|\Omega|^2}{2\bar{\Omega}^2}\cos(\mathbf{k}_L\cdot\mathbf{r}_{12})\left[1-2(\rho_{22}+\rho_{33})\right],
\nonumber \\
U^{\pm}(\mathbf{r}_{12})&=&-\hbar\Delta^{\pm}_{12}\frac{(\delta\mp\bar{\Omega})^2}{4\bar{\Omega}^2}\left[e^{\pm i\mathbf{k}_L\cdot\mathbf{r}_{12}}\rho_{23}+e^{\mp i\mathbf{k}_L\cdot\mathbf{r}_{12}}\rho^{\ast}_{23}\right].
\nonumber \\
\label{LIDDI1}
\end{eqnarray}
Here we have separated the potential $U$ into its components $U^i$ according to its contributions from processes mediated by different photon mode frequencies $\omega_i$ ($i=\pm,z$). For example, if $\rho$ does not allow for a superposition of $|\pm\rangle$ states in each atom, then $\rho_{23}$ vanishes and the only contribution to the LIDDI originates from the $\omega_z=\omega_L$ mediated process, later interpreted here as the linear component of LIDDI. We recall that $\Omega$ and $\Delta_{12}$ contain the atomic-dipole matrix element $\mathbf{d}$ which, in spherically-symmetric cases, is determined by the laser polarization $\mathbf{e}_L$, the only preferred direction (see Appendix B).

\subsubsection{Steady-state and transient potential}
The two-atom density matrix $\rho$, to be used in Eq. (\ref{LIDDI1}) for the LIDDI potential, is determined by the solution of the master equation (\ref{ME2}). In general, this yields a time-dependent potential $U(\mathbf{r}_{12})$ obtained by averaging in Eq. (\ref{LIDDI}) with the time-dependent solution $\rho(t)$. If however the time-resolution of interest is longer than the typical decay times $\tau_z=[\Gamma^z_{\nu\nu'}|\Omega|^2/(2\bar{\Omega}^2)]^{-1}$ and $\tau_{\pm}=[\Gamma^{\pm}_{\nu\nu'}(\delta\mp\bar{\Omega})^2/(4\bar{\Omega}^2)]^{-1}$, then the steady-state solution of the master equation, found by setting $\dot{\rho}=0$ in (\ref{ME2}), can be used to obtain the stationary LIDDI potential. On the contrary, if the dynamics of interest occurs at times much shorter than the decay times $\tau_i$ ($i=\pm,z$), then $\rho(t)$ and hence $U(\mathbf{r}_{12})$, may be obtained by simply solving Eq. (\ref{ME2}) without the dissipative terms. These two limiting cases are illustrated in Sec. IV below for atoms in a cavity.

\subsubsection{Scattering rate}

Turning to the dissipative counterpart of the LIDDI potential, the dressed atoms can scatter laser photons via spontaneous emission, resulting in a random diffusive motion for the atoms, on top of that driven by the LIDDI potential. Upon ignoring (for the time being) the cooperative scattering terms, the scattering rate from a single atom (e.g. atom 1 without loss of generality) can be obtained by averaging the dissipative, imaginary-Hamiltonian-like, terms of the master equation for $\nu=\nu'=1$. In analogy to Eq. (\ref{LIDDI1}), we obtain the single atom scattering rate $R$ as
\begin{eqnarray}
R&=&R^z+R^++R^-,
\nonumber \\
R^z&=&\Gamma^z_{11}\frac{|\Omega|^2}{2\bar{\Omega}^2},
\nonumber \\
R^{\pm}&=&\Gamma^{\pm}_{11}\frac{(\delta\mp\bar{\Omega})^2}{4\bar{\Omega}^2}\langle \pm|\rho^{(1)} |\pm\rangle,
\label{R}
\end{eqnarray}
where $\rho^{(1)}=\mathrm{tr}_2[\rho]$ is the density operator of atom 1, obtained by tracing over atom 2, and $|\pm\rangle$ are the states of atom 1. Here we used $\mathrm{tr}[\rho^{(1)}(\hat{S}^z_1)^2]=1$ and  $\mathrm{tr}[\rho^{(1)}\hat{S}^{\pm}_1\hat{S}^{\mp}_1]=\langle \pm|\rho^{(1)} |\pm\rangle$.

The scattering rate given above is composed of scattering rates of photons at three distinct frequencies: $\omega_z=\omega_L$, the frequency of the incoming laser light, and two sidebands at frequencies  $\omega_{\pm}=\omega_L\pm \bar{\Omega}$, which reproduce the Mollow-triplet fluorescence spectrum. The amplitudes of the sidebands are determined by their corresponding scattering processes: An atom that scatters a $\omega_{\pm}$ photon is initially in state $|\pm\rangle$ and finally in state $|\mp\rangle$. Hence, the amplitude is related to the probability of the atom to be in the initial state, i.e. $\langle \pm|\rho |\pm\rangle$ that appears in the expressions for $R^{\pm}$. Likewise, scattering at $\omega_L$ does not change the internal state, which explains its independence of the atomic state $\rho$.

\subsection{Linear and nonlinear LIDDI}
Our LIDDI result of Eq. (\ref{LIDDI1}) reveals that LIDDI is mediated by virtual photons centered around three frequencies; namely, that of the incident laser, $\omega_L$, and those of the sidebands $\omega_{\pm}$. The former, linear, process gives rise to the LIDDI term proportional to $\Delta_{12}(\omega_L)$, whereas the latter give rise to a nonlinear LIDDI process manifest by  $\Delta_{12}(\omega_{\pm})$. Here, the term "linear" is more easily understood when we consider the scattering rate $R$ [see Eq. (\ref{R})]: if the scattered real photons are at the same frequency as the incident light, as in the term $\Gamma_{11}(\omega_L)$, then the process is linear, as in elastic scattering. However, if the scattered photons are at a different frequency, here $\omega_{\pm}=\omega_L\pm\bar{\Omega}$, then the scattering is inelastic and we call it a nonlinear process. Likewise, viewing LIDDI as the virtual-photon (and cooperative) counterpart of scattering, we call the process mediated by virtual photons around $\omega_z=\omega_L$ "linear" and that mediated by $\omega_{\pm}$ "nonlinear".
As discussed below, previous treatments of LIDDI \cite{MQED,THI,SAL,AND1,AND2} obtained only the linear contribution, whereas our formalism accounts also for the nonlinear response of the atoms to the laser light.

\section{Large detuning limit}

Most of the previous results of LIDDI, e.g. those of Refs. \cite{MQED,THI,SAL,AND1,AND2}, are obtained in the limit of weak laser amplitude $\Omega$ compared to atom-laser detuning $\delta$, assuming also that the atoms are in the ground state. Na\"{\i}vely, this result can be reproduced by our formalism in this large-detuning limit $\Omega/\delta\ll 1$ as follows: We first note that
\begin{equation}
|g\rangle\approx \left(1-\frac{|\Omega|^2}{8}\right)|+\rangle-\frac{\Omega}{2}|-\rangle,
\label{g}
\end{equation}
and further approximate $|g\rangle\approx|+\rangle$ for each atom, such that $\rho=|+_1+_2\rangle\langle+_1+_2|=|1\rangle\langle1|$. We then have $\rho_{22}=\rho_{33}=\rho_{23}=0$, such that we find that the nonlinear LIDDI vanishes, $U^{\pm}=0$, and only the linear part remains,
\begin{equation}
U(\mathbf{r}_{12})\approx U^z(\mathbf{r}_{12})\approx-\frac{|\Omega|^2}{2\delta^2}\hbar\Delta^z_{12}\cos(\mathbf{k}_L\cdot\mathbf{r}_{12}).
\label{UL}
\end{equation}
By recalling the dynamical polarizability of atoms in their ground-state and in the limit of far detuned laser, $\alpha(\omega_L)\approx -|\mathbf{d}|^2/(\hbar\delta)$, and considering the known solution for the RDDI $\Delta_{12}(\omega_L)$ in free space \cite{MQED,LEH}, we recover the free-space LIDDI result from \cite{MQED,THI,SAL}.

This result fails to capture the nonlinear part of LIDDI which is clearly important beyond the weak laser (i.e. large detuning) approximation. Interestingly enough however, the above result, Eq. (\ref{UL}), appears to be inconsistent even within the $\Omega/\delta\ll 1$ approximation, from two main reasons: (1) The assumption that the atoms always stay in the ground state, which although seems reasonable for  large detuning, is not always true, especially for long times (see Sec. IV B below), and generally the state of the atoms $\rho$ has to be found by solving the master equation (\ref{ME2}). (2) Even if we assume that the atoms are in the ground state, i.e. $\rho=|g_1g_2\rangle\langle g_1g_2|$,  and take the approximate $|g\rangle$ from (\ref{g}) instead of just taking $|g\rangle\approx|+\rangle$, we find $\rho_{23}\approx |\Omega|^2/(4\delta^2)$. Then,  to second order in $\Omega/\delta$, there exists also a nonlinear term, describing LIDDI mediated at AC-Stark shifted atomic frequency (for e.g. $\delta>0$) $\omega_-\approx \omega_e-|\Omega|^2/(2\delta)$,
\begin{equation}
U^-(\mathbf{r}_{12}) \approx-\frac{|\Omega|^2}{2\delta^2}\hbar\Delta_{12}(\omega_-)\cos(\mathbf{k}_L\cdot\mathbf{r}_{12}),
\label{UNL}
\end{equation}
which is totally missed by previous treatments and Eq. (\ref{UL}).
This is because in previous treatments linearity, namely, that the mediating virtual photons are at the same frequency as the incident laser photons, was imposed by the method of calculation.
In Appendix C, we briefly review two such previous approaches and show how linearity is imposed by them, thus clarifying the reason for the absence of the nonlinear term therein.

In fact, as shown in Sec. IV below, the above results, (\ref{UL}) and (\ref{UNL}), are obtained for LIDDI in the large-detuning and transient regime, for atoms initially in their ground states, whereas in steady-state only the linear part (\ref{UL}) survives. Moreover, Eqs. (\ref{UL}) and (\ref{UNL}) provide an interesting insight into the nature of LIDDI in the large-detuning limit: Looking at the coefficient $|\Omega|^2/(2\delta^2)$ in front of the RDDI rate $\Delta_{12}$ in both equations, and recalling that it is proportional to the probability to excite an initially ground-state atom in this regime, we can interpret LIDDI as the excitation of an atom by the laser at probability  $|\Omega|^2/(2\delta^2)$ thus allowing it to interact with another atom via RDDI at rate $\Delta_{12}$.

\section{Example: LIDDI in a cavity}
Let us illustrate our formalism by considering the LIDDI potential for atoms inside a cavity geometry. Such an analysis is very relevant for current experimental and theoretical research in many-atom systems inside cavities \cite{RDE}. We treat this example in two cases, the first case where dissipation (emission/scattering) is significant such that the relevant LIDDI is that in the steady state, and the second case in the transient regime where dissipation is negligible. For both cases however, we first have to find the underlying effects of RDDI and emission.

\subsection{RDDI and emission rate}
For simplicity, we assume a perfect cavity of length $L$ and effective area $A$, and consider a single transverse mode, such that the relevant photon modes are those in the longitudinal direction $z$,
\begin{equation}
\mathbf{u}_{n,\mu}(z)=\sqrt{\frac{2}{LA}}\sin\left(\frac{n\pi}{L} z\right)\mathbf{e}_\mu, \quad n=0,1,2,...
\label{un}
\end{equation}
with frequencies $\omega_{n}=n(\pi/L)c$ and where $\mu$ is the polarization index, e.g. $\mu=x,y$.
Assuming all three $\omega_i$'s ($i=\pm,z$) are between and do not coincide with the cavity modes $\omega_n$, there exists no atom dissipation to the cavity modes, i.e. they do not contribute to  $\Gamma_{\nu\nu'}^i$, whereas their contribution to the RDDI, given by Eqs. (\ref{D}), (\ref{un}), is
\begin{equation}
\Delta^i_{12}=\sum_n \frac{\omega_n d_{\bot}^2}{\epsilon_0\hbar L A}\sin\left(n\frac{\pi}{L} z_1\right)\sin\left(\frac{n\pi}{L} z_2\right)\frac{1}{\omega_n-\omega_i},
\label{Dc}
\end{equation}
with $d_{\bot}^2=(\mathbf{d}\cdot\mathbf{e}_x)^2+(\mathbf{d}\cdot\mathbf{e}_y)^2$ (For a more complete treatment of RDDI in a cavity see \cite{SEK}).
Since the transverse ($x,y$) area of the cavity mirrors is finite, the atoms are also coupled to non-confined photon modes. We then expect that their spatial scaling is similar to that of free-space modes, such that their contribution to the cooperative effects $\Gamma^i_{12}$ and $\Delta^i_{12}$ falls off initially like $1/z_{12}^3$ and is negligible for typical atomic distances larger than the atomic wavelength. Then these modes only contribute to the single-atom spontaneous emission rate, which is approximated as that of free-space. Finally, we have
\begin{equation}
\Gamma_{12}^i\approx 0 \quad, \Gamma_{11}^i \approx\Gamma_{fs}(\omega_i), \quad \Gamma_{fs}(\omega)=\frac{ |\mathbf{d}|^2 \omega^3}{3\pi\epsilon_0 \hbar c^3},
\label{Gfs}
\end{equation}
and $\Delta_{12}^i$ from Eq. (\ref{Dc}).

\subsection{Case 1: Steady state LIDDI}
Assuming the dissipation rate is larger or comparable to the  RDDI rate, i.e. $\Gamma_{11}^i=\Gamma_{fs}(\omega_i)\gtrsim \Delta_{12}^i$, there are no interesting dynamics at time-scales much shorter than $[\Gamma_{fs}(\omega_i)]^{-1}$, such that dissipation is important. Then, the relevant LIDDI is that in the steady-state, which is calculated by solving the master equation (\ref{ME2}) for $\dot{\rho}=0$ with the dissipation and RDDI parameters from Eqs. (\ref{Dc}) and (\ref{Gfs}). For the density matrix elements of interest, we find the steady-state equation,
\begin{widetext}
\begin{equation}
\left(
  \begin{array}{ccccc}
    -2\tilde{\Gamma}^+ & \tilde{\Gamma}^- & \tilde{\Gamma}^- & 0 & 0 \\
    -\tilde{\Gamma}^- + \tilde{\Gamma}^+ & -2\tilde{\Gamma}^- - \tilde{\Gamma}^+ & -\tilde{\Gamma}^- & 0 & 2(\tilde{\Delta}^-_{12}+\tilde{\Delta}^+_{12}) \\
    -\tilde{\Gamma}^- + \tilde{\Gamma}^+ & -\tilde{\Gamma}^- & -2\tilde{\Gamma}^- - \tilde{\Gamma}^+ & 0 & -2(\tilde{\Delta}^-_{12}+\tilde{\Delta}^+_{12}) \\
    0 & 0 & 0 &  -\tilde{\Gamma}^--\tilde{\Gamma}^+-4\tilde{\Gamma}^z & 0\\
    0 & -\tilde{\Delta}^-_{12}-\tilde{\Delta}^+_{12} & \tilde{\Delta}^-_{12}+\tilde{\Delta}^+_{12} & 0 & -\tilde{\Gamma}^--\tilde{\Gamma}^+-4\tilde{\Gamma}^z \\
  \end{array}
\right)
\left(
  \begin{array}{c}
    \rho_{11} \\
    \rho_{22} \\
    \rho_{33} \\
    \mathrm{Re}[\rho_{23}] \\
    \mathrm{Im}[\rho_{23}] \\
  \end{array}
\right)
=
\left(
  \begin{array}{c}
    0 \\
   -\tilde{\Gamma}^- \\
     -\tilde{\Gamma}^- \\
    0 \\
    0\\
  \end{array}
\right),
\label{MEst}
\end{equation}
\end{widetext}
with $\tilde{\Gamma}^{\pm}=\Gamma_{11}^{\pm}(\delta\mp\bar{\Omega})^2/(2\bar{\Omega})^2$, $\tilde{\Gamma}^z=\Gamma_{11}^{z}|\Omega|^2/(2\bar{\Omega})^2$ and $\tilde{\Delta}^{\pm}_{12}=\Delta_{12}^{\pm}(\delta\mp\bar{\Omega})^2/(2\bar{\Omega})^2$.
The steady-state solution is then $\rho_{22}=\rho_{33}=\tilde{\Gamma}^-\tilde{\Gamma}^+/(\tilde{\Gamma}^-+\tilde{\Gamma}^+)^2$ and $\rho_{23}=0$. Inserting this solution into the state-dependent LIDDI potential in Eq. (\ref{LIDDI1}) we find only linear LIDDI in steady-state,
\begin{eqnarray}
U(\mathbf{r}_{12})&=&-\frac{|\Omega|^2}{2\bar{\Omega}^2}\left[1-4\frac{\Gamma_{11}^+\Gamma_{11}^-|\Omega|^4}{[\Gamma_{11}^+(\delta-\bar{\Omega})^2+\Gamma_{11}^-(\delta+\bar{\Omega})^2]^2}\right]
\nonumber \\
&&\times \hbar\Delta^z_{12}\cos(\mathbf{k}_L\cdot\mathbf{r}_{12}).
\label{Ust}
\end{eqnarray}
For the large detuning limit, $\Omega/\delta\ll 1$, we expand the above result to lowest order and find,
\begin{equation}
U(\mathbf{r}_{12})\approx-\frac{|\Omega|^2}{2\delta^2}\hbar\Delta^z_{12}\cos(\mathbf{k}_L\cdot\mathbf{r}_{12}),
\label{Ust2}
\end{equation}
reproducing the linear-analysis result shown in Sec. III, Eq. (\ref{UL}).

\subsection{Case 2: Transient LIDDI}
Our analysis shows that while RDDI is mediated by confined cavity photon modes, the dissipation is driven by the non-confined free-space-like modes. For $\omega_L$ (and hence $\omega_i$'s) close to a cavity mode $\omega_n$, this leads to the possibility of RDDI much stronger than dissipation, as revealed by Eqs. (\ref{Dc}) and (\ref{Gfs}). Therefore, in such a regime, it is relevant to consider the dynamics, and hence LIDDI, at time-scales much shorter than the dissipation time $\sim[\Gamma_{fs}(\omega_L)]^{-1}$. This allows to set $\Gamma_{\nu\nu'}^i=0$ in the master equation (\ref{ME2}) for $\rho$, obtaining,
\begin{equation}
\frac{d}{dt}
\left(
  \begin{array}{c}
    \rho_{22} \\
    \rho_{33} \\
    \mathrm{Im}[\rho_{23}] \\
  \end{array}
\right)
=
\left(
  \begin{array}{ccc}
    0 & 0 & 2\Delta \\
    0 & 0 & -2\Delta \\
    -\Delta & \Delta & 0 \\
  \end{array}
\right)
\left(
  \begin{array}{c}
    \rho_{22} \\
    \rho_{33} \\
    \mathrm{Im}[\rho_{23}] \\
  \end{array}
\right),
\label{MEt}
\end{equation}
and $\mathrm{Re}[\dot{\rho_{23}}]=0$, where $\Delta=\tilde{\Delta}^+_{12}+\tilde{\Delta}^-_{12}$. Taking the initial state where both atoms are in the ground state, $\rho(0)=|g_1g_2\rangle\langle g_1g_2|$, we find $\rho_{22}(t)=\rho_{33}(t)=\rho_{23}(t)=\rho_{23}(0)=|\Omega|^2/(4\bar{\Omega}^2)$. The LIDDI potential in the transient regime $t\ll [\Gamma_{fs}(\omega_L)]^{-1}$ then becomes
\begin{eqnarray}
&&U(\mathbf{r}_{12})=U_L(\mathbf{r}_{12})+U_{NL}(\mathbf{r}_{12}),
\nonumber \\
&&U_L(\mathbf{r}_{12})=-\frac{|\Omega|^2\delta^2}{(|\Omega|^2+\delta^2)^2}\hbar\Delta^z_{12}\cos(\mathbf{k}_L\cdot\mathbf{r}_{12}),
\nonumber \\
&&U_{NL}(\mathbf{r}_{12})=-\frac{|\Omega|^2\delta^2}{8(|\Omega|^2+\delta^2)^2}\cos(\mathbf{k}_L\cdot\mathbf{r}_{12})\times
\nonumber \\
&&
\hbar\left[\left(2+\frac{|\Omega|^2}{\delta^2}\right)(\Delta_{12}^-+\Delta_{12}^+)+2\sqrt{1+\frac{|\Omega|^2}{\delta^2}}(\Delta_{12}^--\Delta_{12}^+)\right].
\nonumber\\
\label{Ut}
\end{eqnarray}
Here we divided the LIDDI into its linear and nonlinear contributions, $U_{L}$ and $U_{NL}$, respectively. In the large-detuning approximation $\Omega/\delta\ll 1$ we find
 \begin{eqnarray}
&&U_L(\mathbf{r}_{12})\approx-\frac{|\Omega|^2}{2\delta^2}\hbar\Delta^z_{12}\cos(\mathbf{k}_L\cdot\mathbf{r}_{12}),
\nonumber \\
&&U_{NL}(\mathbf{r}_{12})\approx-\frac{|\Omega|^2}{2\delta^2}\hbar\cos(\mathbf{k}_L\cdot\mathbf{r}_{12})
\times\left\{\begin{array}{c}
\Delta_{12}^- \quad;\quad \delta>0 \\
\Delta_{12}^+ \quad;\quad \delta<0
\end{array}\right..
\nonumber \\
\label{Ut2}
\end{eqnarray}
While the linear part of this result is the same as that of the steady-state result, here there is an additional non-linear contribution, which in the large-detuning regime is equivalent to that discussed in Sec. III, Eq. (\ref{UNL}).

\section{Discussion: Origins of nonlinearity}
Since previous treatments \cite{MQED,THI,SAL,AND1,AND2} have only yielded linear LIDDI processes, it is interesting to focus on the mechanism that leads to nonlinear effects in our more general approach. These effects are manifest in the probing of $\Delta_{12}(\omega)$ at frequencies different than that of the incident laser $\omega_L$ in the expression for LIDDI, Eq. (\ref{LIDDI1}), which is relevant at time resolution $T\gg \bar{\Omega}^{-1}$, with $\bar{\Omega}=\sqrt{|\Omega|^2+\delta^2}$, as revealed by our dynamic master-equation-based formulation (Sec. II B).
In what follows, we discuss, from different perspectives, the origin of the nonlinear effects and when they are expected to be significant.

\subsection{Amplitude modulation of excitation probability}
First, let us take an intuitive dynamic approach towards the fluorescence of an illuminated atom described by Hamiltonian (\ref{HA}). The solution for the dynamics of the atomic excitation-probability $P_e$ is known to be \cite{CCT}
\begin{equation}
P_e(t)=\frac{|\Omega|^2}{\bar{\Omega}^2}  \sin^2\left(\frac{\bar{\Omega}}{2} t\right)=\frac{|\Omega|^2}{2\bar{\Omega}^2}\left[1-\cos(\bar{\Omega} t)\right].
\end{equation}
Hence, the excitation probability $P_e$, and consequently the scattering rate from a single atom, is modulated. Since this scattered radiation is centered around $\omega_L$ [recalling that Hamiltonian (\ref{HA}) is written in a frame rotated by $\omega_L$], this expression implies an amplitude-modulated scattered signal, whose spectrum contains peaks around $\omega_z=\omega_L$ and $\omega_{\pm}=\omega_L\pm\bar{\Omega}$, as found above. The distance between the peaks is $\bar{\Omega}$, hence they can be resolved only at times larger than $\bar{\Omega}^{-1}$. This provides an intuitive interpretation for the scattering rate in Eq. (\ref{R}) and the LIDDI related to it, Eq. (\ref{LIDDI1}): at times $T\gg \bar{\Omega}^{-1}$ it is possible to distinguish between scattered or LIDDI-mediating photons at three frequencies $\omega_i$ ($i=\pm,z$); hence, $\Gamma_{11}(\omega)$ or $\Delta_{12}(\omega)$, respectively, are probed at these frequencies for such a time-resolution.

\subsection{Dressed atoms as modulated open systems}
Let us recall the master-equation approach used to derive the LIDDI and scattering rate expressions above: In Sec. II A we first solved for the relevant operator of the combined atom-laser, or dressed-atom, system that couples it to the reservoir, i.e. $\tilde{\sigma}^{\pm}$. In analogy to the bare-atom case, where $\tilde{\sigma}^{\pm}=\hat{\sigma}^{\pm}e^{\pm i \omega_e t}$, we then proceeded to calculate the RDDI between atoms, however this time they are dressed, with the time-dependent  $\hat{\Sigma}_{\pm}(t)=\tilde{\sigma}_{\pm}e^{\mp i \omega_e t}$ replacing the bare time-independent $\hat{\sigma}_{\pm}$. Then, from the point of view of the reservoir, it interacts with a dipole that oscillates at all frequencies contained in $\hat{\Sigma}_{\pm}(t)$ (around $\omega_e$), and hence responds resonantly with all these frequencies, which are the same $\omega_i$ ($i=\pm,z$) as before. This explains why  $\Gamma_{11}(\omega)$ or $\Delta_{12}(\omega)$ are probed at these (resonant) frequencies. Moreover, it provides a frequency-domain picture for the time-scales at which the nonlinear effect, namely, the probing of $\Gamma_{11}(\omega)$ and $\Delta_{12}(\omega)$ at frequencies other than $\omega_z=\omega_L$, becomes significant: The overlap integral in Eq. (\ref{A1f}) of Appendix A gives distinct results for different $\omega_i$ ($i=\pm,z$) when the sinc function $\delta_T(\omega)$ is narrower than the differences between different $\omega_i$, $T\gg \bar{\Omega}^{-1}$ and as long as the widths of $\Gamma_{11}(\omega)$ and $\Delta_{12}(\omega)$ around $\omega_i$, $1/\tau_c$, are not larger than their difference, $1/\tau_c\lesssim \bar{\Omega}$.

This approach to our results is related to known theories for the interaction between modulated systems and a reservoir \cite{KOF,TLA}: There, modulation of system parameters, such as atomic energy levels, by external fields, yields the replacement of the bare $\hat{\sigma}^{+}$ operator by $\varepsilon(t)\hat{\sigma}^{+}$ in the system-reservoir coupling, $\varepsilon(t)$ being a time-dependent function determined by the modulation, in analogy to $\hat{\Sigma}^{+}(t)$ in our case. By writing $\varepsilon(t)\equiv\sum_q \varepsilon_q e^{i\nu_q t}$, one then obtains that the reservoir response, e.g. $\Gamma_{11}(\omega)$, becomes $\sum_q|\varepsilon_q|^2\Gamma_{11}(\omega_e+\nu_q)$ instead of the bare-system $\Gamma_{11}(\omega_e)$ \cite{KOF,TLA}, in analogy with our LIDDI and bare-atom-RDDI results, respectively.

\subsection{Transition amplitudes between dressed states}
\begin{figure}
\begin{center}
\includegraphics[scale=0.45]{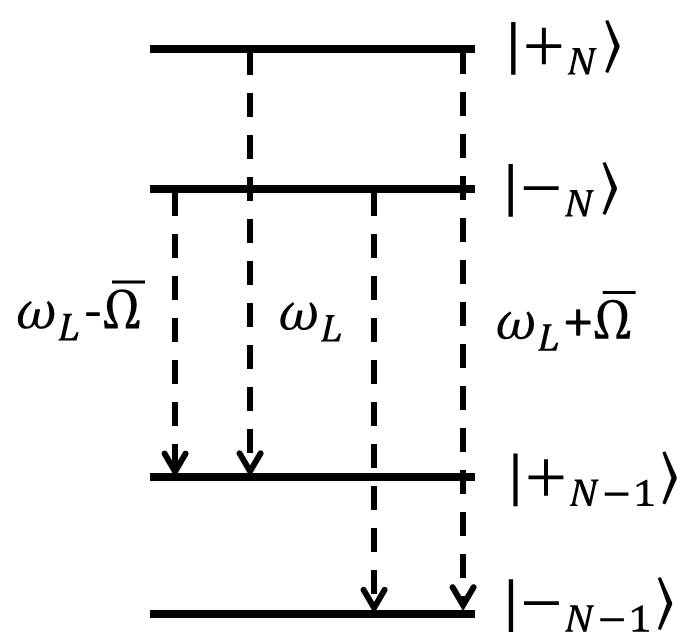}
\caption{\small{
Level diagram of two adjacent manifolds, $N$ and $N-1$, of the dressed atom states, Eq. (\ref{drs}). The possible radiative transitions and their corresponding resonant frequencies are marked with dashed arrows.
}}
\end{center}
\end{figure}

In Section II C we have treated the effects that underlie LIDDI and scattering as dipolar interactions driven by three distinct processes, each involving a different transition between the dressed states of Fig. 1(b). Equations (\ref{HDD_D}), (\ref{LIDDI1}) show that each of these processes has a different amplitude: $|\Omega|^2/(2\bar{\Omega}^2)$ for the process mediated by $\omega_z=\omega_L$ and $(\delta\mp\bar{\Omega})^2/(4\bar{\Omega}^2)$ for the process mediated by $\omega_{\pm}=\omega_L\pm\bar{\Omega}$. In the following we show that these amplitudes are directly related to those of corresponding transitions between laser-quantized dressed states \cite{CCT}.

In the spirit of Chapter VI in Ref. \cite{CCT},  we consider the coupling of a quantized laser mode $\hat{a}$ to an atom via $H=\hbar \omega_L\hat{a}^{\dag}\hat{a}+(1/2)\hbar\omega_e\hat{\sigma}^z+\hbar g_L(\hat{a}^{\dag}\hat{\sigma}^{-}+\hat{a}\hat{\sigma}^+)$ and assume the laser to be in a coherent state with average photon occupation $N$.
For large enough $N$, we can ignore the laser photon fluctuations around the average $\sqrt{N}$ compared to the average $N$, such that in the $N$-manifold of the atom-laser space, $\{|e,N\rangle,|g,N+1\rangle\}$, the relevant Rabi frequency is $\Omega=2g_L\sqrt{N+1}\approx 2g_L\sqrt{N}$  and the corresponding dressed states are
\begin{eqnarray}
|+_N\rangle&=&\frac{(\delta+\bar{\Omega})}{\sqrt{2\bar{\Omega}(\bar{\Omega}+\delta)}}|g,N+1\rangle + \frac{\Omega}{\sqrt{2\bar{\Omega}(\bar{\Omega}+\delta)}}|e,N\rangle,
\nonumber \\
|-_N\rangle&=&\frac{(\delta-\bar{\Omega})}{\sqrt{2\bar{\Omega}(\bar{\Omega}-\delta)}}|g,N+1\rangle + \frac{\Omega}{\sqrt{2\bar{\Omega}(\bar{\Omega}-\delta)}}|e,N\rangle,
\nonumber\\
\label{drs}
\end{eqnarray}
with eigenenergies $(N+1)\hbar\omega_L\pm\hbar \bar{\Omega}$, respectively. The energy level diagram of two adjacent manifolds is plotted in Fig. 2.

Radiative transitions, i.e. emission of a photon from the atom to the vacuum, do not change the number of laser photons but take the atom from the excited to the ground state $|e\rangle\rightarrow |g\rangle$, i.e, involve only a $\hat{\sigma}^-$ operation. Hence, such transitions can only occur between adjacent manifolds. There are four possible transitions, marked by dashed lines in Fig. 2: The two at the center represent the linear process which involves an emission of a photon at the laser frequency $\omega_z=\omega_L$ and the other two transitions at frequencies $\omega_{\pm}=\omega_L\pm\bar{\Omega}$ represent the nonlinear processes.
Hence, the strength of any process that involves an emission or exchange of a photon at frequency $\omega_i$ ($i=\pm,z$) as in the scattering and LIDDI of Eqs. (\ref{R}) and (\ref{LIDDI1}) should be proportional to the probability of the transition between dressed states with energy difference  $\hbar \omega_i$ in the emitting atom.

Figure 2 shows that a nonlinear process with a $\omega_+=\omega_L+\bar{\Omega}$ photon involves the transition from $|+_N\rangle$ to $|-_{N-1}\rangle$ with transition amplitude $\langle -_{N-1}|\hat{\sigma}^-|+_N \rangle=(\delta-\bar{\Omega})/(2\bar{\Omega})$ [inferred from Eq. (\ref{drs})]. The associated probability is then  $|\langle -_{N-1}|\hat{\sigma}^-|+_N \rangle|^2=(\delta-\bar{\Omega})^2/(4\bar{\Omega}^2)$, just as in the prefactor that appears in the $i=+$ scattering and LIDDI of Eqs. (\ref{R}) and (\ref{LIDDI1}). Likewise, the relevant nonlinear $\omega_-=\omega_L+\bar{\Omega}$ process seen in Fig. 2 is that of the $|-_N\rangle\rightarrow |+_{N-1}\rangle$ transition, with amplitude $\langle +_{N-1}|\hat{\sigma}^-| -_N\rangle=(\delta+\bar{\Omega})/(2\bar{\Omega})$ and probability $(\delta+\bar{\Omega})^2/(4\bar{\Omega}^2)$, as in the prefactor of the $i=-$ terms of Eqs. (\ref{LIDDI1}),(\ref{R}). Finally, there are two different processes that involve a $\omega_L$ photon, i.e. $|+_N\rangle\rightarrow |+_{N-1}\rangle$ and $|-_N\rangle\rightarrow |-_{N-1}\rangle$ with identical transition probabilities $|\Omega|^2/(4\bar{\Omega}^2)$. Summing the probabilities of both processes we obtain $|\Omega|^2/(2\bar{\Omega}^2)$ as in the prefactor of the $i=z$ terms of Eqs. (\ref{LIDDI1}),(\ref{R}).

\section{Conclusions}
This study has addressed laser-induced interactions between polarizable dipoles (LIDDI), e.g. atoms, molecules or nanoparticles, in a general geometry. It is expected to prove useful to the understanding and design of LIDDI between polarizable dipoles in confined electromagnetic environments, encountered in various experimental systems \cite{KIM,RAU,RDE,GAE,HAR,LTT,TRA}.

We have analyzed in detail LIDDI in a general geometry without assuming that the dipoles respond linearly to the applied laser field, and have found that it can be described as the resonant dipole-dipole interaction (RDDI) between laser-dressed atoms in the same geometry. This description has allowed us to obtain general formulae for both LIDDI and scattering in terms of the corresponding RDDI and spontaneous emission, respectively [Eqs. (\ref{LIDDI1}) and (\ref{R})].

The LIDDI and scattering revealed by our formulae contain contributions due to the \emph{nonlinear} response of the dipoles to the laser light, which are missing in previous treatments, e.g. those of Ref. \cite{MQED,THI,SAL,AND1,AND2}. By reviewing the principles of these previous treatments, we have demonstrated their imposed linearity as the source of the absence of nonlinear LIDDI/scattering effects in these treatments. We have then explained, using several simple and intuitive approaches, the origin of the nonlinear LIDDI and scattering terms in our expressions.

Considerable progress has been reported in the ability to couple atoms to cavities \cite{RDE,LTT} and nano-fibers \cite{KIM,RAU}. This promising direction has already lead to exciting predictions \cite{RDE,WOLF,CHA,LIDDI} and experiments \cite{RDE,ESS} concerning the many-body physics of illuminated atoms in confined geometries. In this context, the generality of our formalism and its illustration for the analysis of LIDDI in a cavity, suggest its applicability for further exploration of this timely line of research.

\acknowledgements
We appreciate useful discussions with Ilya Averbukh. The support of ISF and BSF is acknowledged.

\appendix

\section{Master equation derivation}
Here we present the derivation of the master equation (ME), Eq. (\ref{ME2}). Beginning with the double commutator in the ME (\ref{ME1}),
$(1/\hbar^2)[H_I(t),[H_I(t'),\rho(t')\rho_V]]$, it includes four terms,
\begin{eqnarray}
&&\frac{1}{\hbar^2}\left[H_I(t)H_I(t')\rho(t')\rho_V-H_I(t)\rho(t')\rho_V H_I(t')
\right.
\nonumber \\
&& \left.
-H_I(t')\rho(t')\rho_V H_I(t)+\rho(t')\rho_V H_I(t')H_I(t)\right].
\label{com}
\end{eqnarray}
The rest of the derivation is similar for these four terms, hence in the following we present in detail only that of the first term. Inserting $H_I(t)$ from Eq. (\ref{HI2}) into the first term in (\ref{com}) we obtain
\begin{eqnarray}
&&\sum_{\nu \nu'}\sum_{ij}\sum_{kk'} \left[
\right. \nonumber\\ && \left.
-g_{k\nu} g_{k'\nu'}\tilde{S}_{i\nu}\tilde{S}_{j\nu'}e^{-i(\omega_k-\omega_i)t}e^{-i(\omega_k'-\omega_j)t'}\rho(t')\hat{a}_k\hat{a}_{k'}\rho_V
\right. \nonumber\\ && \left.
+g_{k\nu}^{\ast} g_{k'\nu'}\tilde{S}_{i\nu}^{\dag}\tilde{S}_{j\nu'}e^{i(\omega_k-\omega_i)t}e^{-i(\omega_k'-\omega_j)t'}\rho(t')\hat{a}_k^{\dag} \hat{a}_{k'}\rho_V
\right. \nonumber\\ && \left.
+g_{k\nu} g_{k'\nu'}^{\ast}\tilde{S}_{i\nu}\tilde{S}_{j\nu'}^{\dag}e^{-i(\omega_k-\omega_i)t}e^{i(\omega_k'-\omega_j)t'}\rho(t')\hat{a}_k \hat{a}_{k'}^{\dag}\rho_V
\right. \nonumber\\ && \left.
-g_{k\nu}^{\ast} g_{k'\nu'}^{\ast}\tilde{S}_{i\nu}^{\dag}\tilde{S}_{j\nu'}^{\dag}e^{i(\omega_k-\omega_i)t}e^{i(\omega_k'-\omega_j)t'}\rho(t')\hat{a}_k^{\dag} \hat{a}_{k'}^{\dag}\rho_V
\right].
\nonumber\\
\label{A1a}
\end{eqnarray}
Taking the trace $\mathrm{tr}_V$ in Eq. (\ref{ME1}) over the reservoir in the vacuum state, we have
 \begin{equation}
\mathrm{tr}_V[\rho_V\hat{a}_k \hat{a}_{k'}^{\dag}]=\langle \hat{a}_k \hat{a}_{k'}^{\dag} \rangle=\delta_{kk'},
\label{tr}
\end{equation}
and similarly $\mathrm{tr}_V[\rho_V\hat{a}_k \hat{a}_{k'}]=\mathrm{tr}_V[\rho_V\hat{a}_k^{\dag} \hat{a}_{k'}]=\mathrm{tr}_V[\rho_V\hat{a}_k^{\dag} \hat{a}_{k'}^{\dag}]=0$, such that only the third term in (\ref{A1a}) contributes, and the trace over (\ref{A1a}) becomes
\begin{equation}
\sum_{\nu \nu'}\sum_{ij}\sum_{k}e^{-i(\omega_k-\omega_i)(t-t')}g_{k\nu} g_{k\nu'}^{\ast}\tilde{S}_{i\nu}\tilde{S}_{j\nu'}^{\dag}\rho(t')e^{i(\omega_i-\omega_j)t'}.
\label{A1b}
\end{equation}
Assuming that the relevant time-scale for the effects we are interested to resolve (coarse graining resolution), $T$, is much larger than $1/|\omega_i-\omega_j|$ for any $i\neq j$, or simply that $T\gg \bar{\Omega}^{-1}=1/\sqrt{|\Omega|^2+\delta^2}$, we can view the exponent on the right hand side of (\ref{A1b}) $e^{i(\omega_i-\omega_j)t'}$ as a fast oscillation that averages out for $i\neq j$ and hence replace it with a Kronecker delta $\delta_{ij}$, yielding for (\ref{A1b})
\begin{equation}
\sum_{\nu \nu'}\sum_{i}\sum_{k}e^{-i(\omega_k-\omega_i)(t-t')}g_{k\nu} g_{k\nu'}^{\ast}\tilde{S}_{i\nu}\tilde{S}_{i\nu'}^{\dag}\rho(t').
\label{A1c}
\end{equation}
Taking the continuum limit $\sum_k g_{k\nu}g_{k\nu'}^{\ast}\longrightarrow \int d\omega D(\omega) g_{\nu}(\omega)g^{\ast}_{\nu'}(\omega)$, where $D(\omega)$ is the density of photon modes $\{k\}$ in the considered geometry [e.g. $D(\omega)\propto\omega^2$ in free-space or $D(\omega)\propto\partial\beta/\partial\omega$ in a waveguide with on-axis wavenumber $\beta$], we identify the vacuum reservoir two-point (autocorrelation) spectrum as
\begin{equation}
G_{\nu\nu'}(\omega)=D(\omega) g_{\nu}(\omega)g^{\ast}_{\nu'}(\omega).
\label{G}
\end{equation}
Then, performing the integration $-\int_0^t dt'$ in the ME (\ref{ME1}) and using (\ref{G}), the term in (\ref{A1c}) becomes
\begin{equation}
-\sum_{\nu \nu'}\sum_{i}\int d\omega G_{\nu\nu'}(\omega) \int_0^t dt' e^{-i(\omega-\omega_i)(t-t')}\tilde{S}_{i\nu}\tilde{S}_{i\nu'}^{\dag}\rho(t'),
\label{A1d}
\end{equation}
such that the ME has $\dot{\rho}(t)$ in its left-hand side (LHS) and the above term (\ref{A1d}), along with 4 similar terms, in the right-hand side (RHS).

\emph{Markov approximation}. Assume now that we are interested to solve the ME for very short times $t=T$, much shorter than the typical time-scale of variation for $\rho(t)$. Namely, we assume that $\rho(t')$ hardly changes within the integral from $0$ to $t$, such that it can be taken out of the integral and approximated as $\rho(t')\approx\rho(T)$, which is the so-called Markov approximation. At first, this seems like a strange assumption, considering the fact that eventually we are interested to find an equation for the dynamics of $\rho(t)$ on time-scales $t$ where $\rho(t)$ does change appreciably, but this will become clearer in the following, where $T$ is interpreted as a coarse-graining time. Next, we need to solve the resulting integral over $dt'$,
\begin{equation}
\int_0^T dt' e^{-i(\omega-\omega_i)(T-t')}=\int_{-T}^{T} d\tau\Theta(\tau)e^{-i(\omega-\omega_i)\tau},
\label{int}
\end{equation}
where the coordinate transformation $\tau=T-t'$ was used and $\Theta(x)$ is the Heaviside step function. Using also the relations
\begin{eqnarray}
&&\Theta(\tau) e^{-i(\omega-\omega_i)\tau}=-\lim_{\eta\rightarrow 0^+} \frac{1}{2\pi i}\int_{-\infty}^{\infty}d\omega'\frac{e^{-i\omega' \tau}}{\omega'+i\eta-(\omega-\omega_i)},
\nonumber \\
&&\lim_{\eta\rightarrow 0^+} \frac{1}{\omega'+i\eta-(\omega-\omega_i)}
\nonumber \\
&&=i\pi\delta(\omega-(\omega'+\omega_i))+\mathrm{P}\frac{1}{\omega-(\omega'+\omega_i)},
\label{th}
\end{eqnarray}
where $\mathrm{P}$ denotes the principal value upon integration, the ME term (\ref{A1d}) becomes
\begin{eqnarray}
&-&\sum_{\nu \nu'}\sum_{i}\tilde{S}_{i\nu}\tilde{S}_{i\nu'}^{\dag}\rho(T)\int_{-\infty}^{\infty}d\omega'  \int_{-T}^{T} d\tau
e^{-i\omega'\tau}\int d\omega G_{\nu\nu'}(\omega)
\nonumber \\
&&\times \left[-\frac{1}{2}\delta(\omega-(\omega'+\omega_i))+\frac{1}{2\pi}i\mathrm{P}\frac{1}{\omega-(\omega'+\omega_i)}\right].
\label{A1e}
\end{eqnarray}
Denoting the sinc function $\delta_T(\omega)\equiv 1/(2\pi)\int_{-T}^Td\tau e^{-i\omega\tau}$, (\ref{A1e}) becomes
\begin{eqnarray}
&&\sum_{\nu \nu'}\sum_{i}\tilde{S}_{i\nu}\tilde{S}_{i\nu'}^{\dag}\rho(T)
\nonumber \\
&&\times \int_{-\infty}^{\infty}d\omega' \delta_T(\omega')\left[\frac{1}{2}\Gamma_{\nu\nu'}(\omega'+\omega_i)-i\Delta_{\nu\nu'}(\omega'+\omega_i)\right],
\nonumber\\
\label{A1f}
\end{eqnarray}
with $\Gamma_{\nu\nu'}(\omega)$ and $\Delta_{\nu\nu'}(\omega)$ from Eq. (\ref{D}) in the main text. This expression describes an overlap integral between the complex linear response (spectrum) of the vacuum reservoir to the two-atom system $(1/2)\Gamma_{\nu\nu'}(\omega)+i\Delta_{\nu\nu'}(\omega)$ and the sinc function  $\delta_T(\omega)$ of width $\sim 1/T$ around $\omega=0$. Denoting the typical width of the vacuum spectral response functions $\Gamma_{\nu\nu'}(\omega)$ and $\Delta_{\nu\nu'}(\omega)$ around $\omega_i$ by $1/\tau_c$, and assuming $T\gg \tau_c$, we can approximate $\delta_T(\omega)\rightarrow \delta(\omega)$ and finally get
\begin{equation}
\sum_{\nu \nu'}\sum_{i}\left[\frac{1}{2}\Gamma_{\nu\nu'}(\omega_i)-i\Delta_{\nu\nu'}(\omega_i)\right]\tilde{S}_{i\nu}\tilde{S}_{i\nu'}^{\dag}\rho(T).
\label{A1g}
\end{equation}
The rest of the three terms that appear on the RHS of the ME can be treated similarly, thus obtaining the ME in Eq. (\ref{ME2}). However, here it is written for short times $t=T$ w.r.t the typical time-variation of $\rho(t)$. Nevertheless, since the coefficients of the ME are constant, it has the form
\begin{equation}
\dot{\rho}(T)=\mathcal{L}\rho(T),
\end{equation}
with $\mathcal{L}$ a time-independent superoperator. Then, its formal solution for $t=T$ is $\rho(T)=e^{\mathcal{L}T}\rho(0)$, and similarly $\rho(2T)=e^{\mathcal{L}T}\rho(T)=e^{\mathcal{L}2T}\rho(0)$ for $t=2T$. This procedure can be repeated successively for any time duration $t$, so that the ME is in fact valid for arbitrary long time $t$, including time durations where the system state $\rho(t)$ changes considerably.

\emph{Separation of time-scales}. The coefficients of the ME terms become time-independent in the Markov approximation, namely by assuming that the time-resolution of interest, $T$, is much larger than  the so-called correlation time of the reservoir $\tau_c$. On the other hand, in order to have sufficient resolution in probing the system dynamics, recall that we assumed $T$ to be much shorter than the typical time of variation for the system, that can be directly read from the ME to be $\sim \Gamma^{-1}$, i.e. the inverse of the rates $\Gamma^i_{\nu\nu'},\Delta^i_{\nu\nu'}$. This implies,
\begin{equation}
\tau_c\ll T \ll \Gamma^{-1},
\end{equation}
namely, that the time-scales for the reservoir memory/correlation, $\tau_c$, and the system dynamics, $\Gamma^{-1}$, are well separated.

\section{LIDDI dependence on laser polarization}
Let us consider the dependence of the LIDDI potential, Eq. (\ref{LIDDI1}), on the orientation of the dipolar matrix element $\mathbf{d}$  and laser polarization. We note that both $\Omega$ and $\Delta_{12}$ (via $g_{k\nu}\propto \mathbf{d}\cdot \mathbf{u}_k$) depend on $\mathbf{d}$, which we recall to be $\mathbf{d}=\langle g| \hat{\mathbf{d}}|e\rangle$, where $\hat{\mathbf{d}}$ is the dipole operator (e.g. the electron charge times electron position operator in an atom).
For the case of polarizable dipoles with a fixed orientation, $\mathbf{d}$ is fixed, and so are the products $\mathbf{d}\cdot\mathbf{u}_k$ and $\mathbf{d}\cdot\mathbf{e}_L$, such that $\Omega$, $\Delta_{12}$ and hence the LIDDI $U$ can be calculated. However, for cases where the atomic polarizability is effectively isotropic, the only preferred direction is that of the laser polarization, such that it effectively determines the orientation of $\mathbf{d}$. In the following we illustrate this idea for two typical scenarios of atomic and molecular dipoles.

\emph{Isotropic atoms}. Consider the ground state $|g\rangle$ to have a spherical symmetric electronic wavefunction, i.e. with spherical harmonic quantum numbers $l,m=0,0$: $|g\rangle=|0,0\rangle$. Then, the quantization axis $z$ that fixes the quantum number $m$ is arbitrary and, assuming the laser is linearly polarized, we can choose it to be that of the laser polarization, i.e. $\mathbf{e}_z=\mathbf{e}_L$. Consider now the possible dipole-allowed excited states in a spherically symmetric potential for the electron, $|1,m\rangle$ with $m=0,\pm 1$. Then, by dipole selection rules, the dipole operator is generally written as
\begin{equation}
\hat{\mathbf{d}}=d_+ \mathbf{e}_+ |0,0\rangle \langle 1,1|+d_- \mathbf{e}_- |0,0\rangle \langle 1,-1|+d_z\mathbf{e}_z |0,0\rangle \langle 1,0|,
\label{d}
\end{equation}
where $\mathbf{e}_{\pm}$ is a right/left circular polarization unit vector, respectively.
Then, the product $\mathbf{d}\cdot\mathbf{e}_L=\mathbf{d}\cdot\mathbf{e}_z$ in $\Omega$ implies that the only state to be excited by the laser is that of $m=0$, namely, $|e\rangle=|1,0\rangle$, such that the relevant dipolar matrix element that appears in the LIDDI potential $U$ is $\mathbf{d}=d_z \mathbf{e}_z=d_z \mathbf{e}_L$.

Considering a circular laser polarization, $\mathbf{e}_L=\mathbf{e}_{\pm}$, the arbitrary quantization axis $z$ is now chosen to be perpendicular to both circular polarizations, and accordingly the product $\mathbf{d}\cdot\mathbf{e}_L$ in $\Omega$ imposes $|e\rangle=|1,\pm 1\rangle$ and $\mathbf{d}=d_{\pm}  \mathbf{e}_{\pm}=d_{\pm} \mathbf{e}_L$, again parallel to the laser polarization.

\emph{Randomly oriented anisotropic molecules}. Consider now molecules with an anisotropic polarizability, e.g. where for a given quantization axis not all possible $m=0,\pm 1$ excited states exist, and the dipole matrix element is some fixed $\mathbf{d}$. However, by allowing for random molecule orientation, the quantization axis of each molecule (with index $\nu$) becomes random and so does the orientation of its corresponding dipole matrix element $\mathbf{d}_{\nu}$. We recall,
\begin{eqnarray}
\Delta_{12}&\propto&\sum_k g_{k1}g^{\ast}_{k2}, \quad g_{k\nu}\propto \mathbf{d}_{\nu}\cdot \mathbf{e}_k,
\quad \Omega_{\nu}\propto \mathbf{d}_{\nu}\cdot\mathbf{e}_L
\nonumber \\
U&\propto& \Delta_{12}\Omega_1\Omega^{\ast}_2\propto  \sum_k(\mathbf{d}_1\cdot \mathbf{e}_k)( \mathbf{d}_2\cdot \mathbf{e}^{\ast}_k)(\mathbf{d}_1\cdot \mathbf{e}_L) (\mathbf{d}_2\cdot \mathbf{e}_L^{\ast}),
\nonumber \\
\label{u12}
\end{eqnarray}
where $\mathbf{e}_k$ is the unit vector of $\mathbf{u}_k$ and where we considered for simplicity $U\propto \Delta_{12}\Omega_1\Omega^{\ast}_2$ in the large-detuning approximation as in Eqs. (\ref{UL}) and (\ref{UNL}). The, choosing the spherical coordinates of the vector $\mathbf{d}_{\nu}$ with angles $\theta_{\nu}$ and $\phi_{\nu}$, setting $\mathbf{e}_z=\mathbf{e}_L$, we have
\begin{eqnarray}
\mathbf{d}_{\nu}\cdot \mathbf{e}_k&=&|\mathbf{d}|[ e_{k}^x\sin\theta_{\nu}\cos\phi_{\nu}+ e_{k}^y\sin\theta_{\nu}\sin\phi_{\nu}
+e_{k}^z\cos\theta_{\nu}]
\nonumber \\
\mathbf{d}_{\nu}\cdot \mathbf{e}_L&=&|\mathbf{d}|\cos(\theta_{\nu}),
\end{eqnarray}
with $e_{k}^i=\mathbf{e}_k\cdot\mathbf{e}_i$ ($i=x,y,z$). Inserting the above products into $U$ in (\ref{u12}), and assuming no correlation between the random orientation of different molecules, we obtain
\begin{eqnarray}
U&\propto& \sum_k \prod_{\nu=1}^2|\mathbf{d}| \left[e_{k}^x\langle \sin\theta_{\nu}\cos\phi_{\nu}\cos\theta_{\nu}\rangle
\right. \nonumber \\
&&\left. +e_{k}^y\langle \sin\theta_{\nu}\sin\phi_{\nu}\cos\theta_{\nu}\rangle+e_{k}^z\langle\cos^2\theta_{\nu}\rangle\right],
\end{eqnarray}
where $\langle...\rangle\equiv1/(4\pi)\int_0^{2\pi}d\phi_{\nu}\int_0^{\pi}d\theta_{\nu}\sin\theta_{\nu}...$ denotes averaging over orientation with an isotropic distribution. Performing the averaging we finally get
\begin{equation}
U\propto\sum_k(1/3)^2 |\mathbf{d}|^2 (e_{k}^z)^2=(1/9)\sum_k|\mathbf{d}|^2(\mathbf{e}_L\cdot\mathbf{e_k})^2.
\end{equation}
This result is equivalent to taking $\mathbf{d}=(1/3)|\mathbf{d}| \mathbf{e}_L$ in $\Delta_{12}$ and $\Omega$. Therefore, again, yet for a different spherically symmetric system, we effectively obtained  $\mathbf{d}\propto \mathbf{e}_L$.

\section{Review of previous linear treatments}
In section III we have mentioned that the apparent discrepancy between our LIDDI results and those obtained by previous treatments (e.g. \cite{MQED,THI,SAL,AND1,AND2}) is due to the linearity imposed in those treatments. Here we would like to briefly review the essence of two typical approaches used previously, and highlight how linearity is imposed in them, thus preventing them to capture the nonlinear terms of LIDDI revealed by our formalism.

\subsection{First approach: Lowest-order scattering}
Consider the following description of LIDDI depicted in Fig. 3(a): a strong laser beam illuminates two atoms, such that atom 1 is subject to the laser electric field $E_L(\mathbf{r}_1)$ and atom 2 to the field $E_L(\mathbf{r}_2)$. Focusing on atom 1, the laser induces its polarization, which to first order in $E_L$ is $P_1=\alpha_1 E_L(\mathbf{r}_1)$, $\alpha_1$ being the atomic linear polarizability at the laser frequency $\omega_L$. Then, since the polarized atom 1 is an oscillating dipole, it scatters a field $E_{sc}$, which then arrives at atom 2 where it is given to lowest order (Born approximation) by $E_{sc}^{(1)}(\mathbf{r}_2)=P_1 K(\omega_L,\mathbf{r}_{12})$. Here $K(\omega_L,\mathbf{r}_{12})$ is proportional to the Green's function of the electromagnetic-field propagation at frequency $\omega_L$ from $\mathbf{r}_1$ to $\mathbf{r}_2$ in the considered geometry. Since the electromagnetic energy of atom 2 (like any dipole) is $-(1/2)\alpha_2 E_2^2$, $E_2$ being the total electric field at $\mathbf{r}_2$, the lowest-order interaction energy $U$, interpreted as LIDDI, becomes,
\begin{equation}
U\propto\alpha_1\alpha_2 E_L(\mathbf{r}_1) E_L(\mathbf{r}_2) K(\omega_L,\mathbf{r}_{12})\propto \frac{|\Omega|^2}{\delta^2}K(\omega_L,\mathbf{r}_{12}),
\label{U1}
\end{equation}
where the last step (right-hand side) is valid for atoms in their ground state and for large detuning. The result and description above render the essence of treatments used in Refs. like \cite{THI,SAL,AND1,AND2}. Its analogy to the linear part of our result (\ref{UL}), $\propto (|\Omega|^2/\delta^2)\Delta_{12}(\omega_L)$, becomes transparent upon recalling that the RDDI term $\Delta_{12}(\omega)$ is directly related to the Green's function $K(\omega,\mathbf{r}_{12})$ \cite{FLE}. The origin of linearity in this treatment is clear: the atoms are modeled as polarizable dipoles by using their linear response to electric fields, $\alpha_{1,2}$, at the same frequency as the incident radiation. This leads to the sampling of the Green's function at frequency $\omega_L$, thereby obtaining $U\propto K(\omega_L,\mathbf{r}_{12}),\Delta_{12}(\omega_L)$. Clearly, by imposing a description where the highly nonlinear two-level atoms linearly respond to the strong laser, effects due to real or virtual photons at frequencies other than $\omega_L$ cannot be revealed.

\begin{figure}
\begin{center}
\includegraphics[scale=0.4]{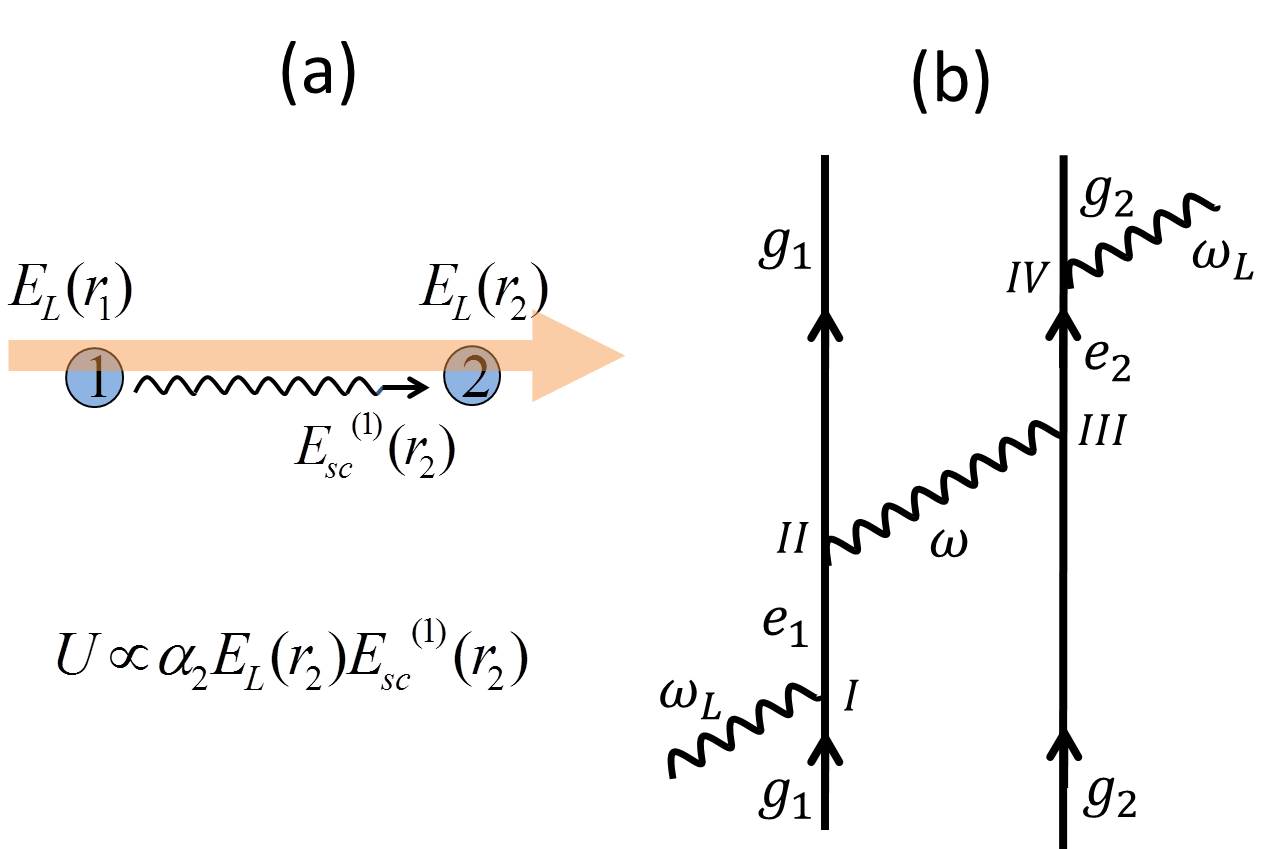}
\caption{\small{
(Color online) Linear approaches for laser-induced dipolar interactions. (a) Dipole 1 responds linearly to the applied laser field $E_L(\mathbf{r}_1)$ and scatters a field $E_{sc}^{(1)}(\mathbf{r}_2)$. Dipole 2 responds linearly to the total field at $\mathbf{r}_2$, which result in the interaction energy $U$, interpreted as the LIDDI potential (see Sec. III A). (b) One of 24 the diagrams that represent sets of intermediate states in the sum of Eq. (\ref{U2}) for the LIDDI potential. The intermediate states corresponding to the above diagram are $|I_1\rangle=|e_1,g_2,0\rangle|N-1\rangle_L$, $|I_2\rangle=|g_1,g_2,1_{\omega}\rangle|N-1\rangle_L$ and $|I_3\rangle=|g_1,e_2,0\rangle|N-1\rangle_L$, where $|1_\omega\rangle\equiv\hat{a}_{\omega}^{\dag}|0\rangle$.
}}
\end{center}
\end{figure}

\subsection{Second approach: Fourth-order QED perturbation theory}
A different approach, applied in e.g. Refs.\cite{MQED,SAL}, uses a QED time-independent perturbative treatment similar to that used to calculate the van der Waals and Casimir-Polder forces. Taking the state of the laser as a number state for the time-being, the laser and the vacuum are treated as a single system (the photons), where there are $N$ photons in the mode with $\omega_L$ and vacuum in the rest of the modes. Then, the LIDDI potential is interpreted as the lowest (fourth) order correction (in the atom-photon coupling) of the state $|\psi\rangle=|g_1,g_2,0\rangle|N\rangle_L$, namely, both atoms in the ground state, $N$ photons in the laser mode $L$ and vacuum ($0$) in the rest of the modes. This energy correction is then given by \cite{MQED,SAL}
\begin{eqnarray}
&&U=-\sum_{I_1,I_2,I_3}
\nonumber\\
&&\frac{\langle \psi|H_{AV}|I_3\rangle \langle I_3|H_{AV}|I_2\rangle \langle I_2|H_{AV}|I_1\rangle \langle I_1|H_{AV}|\psi\rangle}
{(E_{I_1}-E_{\psi})(E_{I_2}-E_{\psi})(E_{I_3}-E_{\psi})},
\nonumber\\
\label{U2}
\end{eqnarray}
with $H_{AV}$ from Eq. (\ref{H}). This is a sum over intermediate/virtual states, $I_1$ ,$I_2$ and $I_3$, where $E_q$ is the energy of state $|q\rangle$ without the interaction $H_{AV}$, e.g. $E_{\psi}=N\hbar \omega_L$ (taking the atomic ground-state energy to zero here). The possible sets of virtual states can be represented by diagrams, such as that in Fig. 3(b), with a total of 24 diagrams \cite{MQED,SAL}. In order to illustrate the essence of this approach, it is enough to focus on the diagram in Fig. 3(b): it describes the intermediate states $|I_1\rangle=|e_1,g_2,0\rangle|N-1\rangle_L$, $|I_2\rangle=|g_1,g_2,1_{\omega}\rangle|N-1\rangle_L$ and $|I_3\rangle=|g_1,e_2,0\rangle|N-1\rangle_L$ with energies $E_{I_1}=(N-1)\hbar\omega_L+\hbar \omega_e$, $E_{I_2}=(N-1)\hbar\omega_L+\hbar \omega$ and $E_{I_3}=(N-1)\hbar\omega_L+\hbar \omega_e$, respectively, where $\omega$ denotes the mode of the virtual photon that mediates the interaction between vertices II and III, with $|1_\omega\rangle\equiv\hat{a}_{\omega}^{\dag}|0\rangle$. By identifying the laser Rabi frequency as $\Omega=g_L\sqrt{N}$, $g_L$ being the dipole coupling to the laser mode, the resulting energy correction for this diagram becomes
\begin{eqnarray}
U&\propto&\int d\omega \frac{\Omega^{\ast} G_{12}(\omega) \Omega}{(\omega_e-\omega_L)(\omega-\omega_L)(\omega_e-\omega_L)}
\nonumber \\
&=&\frac{|\Omega|^2}{\delta^2}\int d\omega \frac{G_{12}(\omega)}{\omega-\omega_L}\propto\frac{|\Omega|^2}{\delta^2}\Delta_{12}(\omega_L),
\label{U3}
\end{eqnarray}
yielding again the linear LIDDI. The crucial step that leads to linearity here is the treatment of atom-laser interaction in vertices I and IV: these vertices give rise to the factor $|\Omega|^2/\delta^2$ which is indeed small for large detuning. However, such a treatment, where the interaction with the strong laser is taken only in two bare atom-laser vertices, does not allow to account for the nonlinear effect.

\end{document}